\documentclass[twocolumn,aps,floatfix,showpacs,10pt,prx]{revtex4-1}
\usepackage{amsmath}
\usepackage{graphicx}
\usepackage{epsfig}
\usepackage{graphics}
\usepackage{bm}
\usepackage{amssymb}
\usepackage{psfrag}
\usepackage[T1]{fontenc}

\begin{document}
\title{Emergent non-trivial lattices for topological insulators}
\author{O. Dutta$^{1,2}$}
\author{A. Przysi\k{e}\.zna$^{3,4}$, and M. Lewenstein$^{1,5}$}
\affiliation{
\mbox{$^1$ ICFO ---Institut de Ciencies Fotoniques, Av. Carl Friedrich Gauss, num. 3, 08860 Castelldefels (Barcelona), Spain  }
\mbox{$^2$ Instytut Fizyki imienia Mariana Smoluchowskiego, Uniwersytet Jagiello{\'n}ski, ulica Reymonta 4, PL-30-059 Krak\'ow, Poland }
\mbox{$^3$ Institute of  Theoretical Physics and Astrophysics, University of Gda\'nsk, Wita Stwosza 57, 80-952 Gda\'nsk, Poland }
\mbox{$^4$ National Quantum Information Centre of Gda\'nsk, Andersa 27, 81-824 Sopot, Poland}
\mbox{$^5$ ICREA -- Instituci{\o} Catalana de Recerca i Estudis Avan\c{c}ats, Lluis Companys 23, E-08010 Barcelona, Spain} }

\date{\today}

\begin{abstract}
{Materials with non-trivial lattice geometries allow for the creation of exotic states of matter like topologically insulating states. Therefore searching for such materials is an important aspect of current research in solid-state physics. In the field of ultracold gases there are ongoing studies aiming to create non-trivial lattices using optical means. In this paper we study two species of
fermions trapped in a square optical lattice and show how non-trivial lattices can emerge due to strong interaction between atoms. We theoretically investigate regimes of tunable parameters in which such self-assembly may take place and describe the necessary experimental conditions. Moreover we discuss the possibility of such emergent lattices hosting topologically insulating states. }
\end{abstract}

\maketitle
{
Complex systems are characterized by large number of locally interacting elements with properties that cannot be derived as a sum of local individual elements \cite{Anderson}. In such systems spontaneous self-assembly \cite{assembly} takes place and {\it emergent}  structures are formed from disorder due to the cooperative effect of the interacting system. Such emergent behaviour of complex systems is responsible for many organized structures found in many-body physics, chemistry, biological systems etc. 
In the present paper we show that cooperation of interaction and orbital effects in a lattice can result in non-trivial emergent structures(lattices) with topological order.

Non-trivial lattice geometries are at the heart of various exotic
phenomena in many-body physics. One promising playground to realize such exotic lattices and states are ultracold gases trapped in lattice potentials \cite {Lew}. This is due to high degree of experimental controllability and tunability that ultracold systems exhibit. In this field, variety of nontrivial lattice geometries were experimentally created by using 
counter-propagating laser beams in different configurations \cite{Seng, Stamp, Hemm, Ess}. Such lattices can be than used to realize exotic states by introducing tunable long range hopping amplitudes \cite{Bloch, Lew1, Liu, Mudry, Gorshkov}. Particularly interesting is the  possibility of creating topologically insulating states \cite{Hasan,
Zhang1, Haldane} that allow for a robust transport of charges
(matter) on the boundary and thus have potential applications in
spintronics, quantum computing \cite{Nayak} and spintomics \cite{Lew}.

In this paper we propose an alternative route to create non-trivial lattice geometries. Our system consists of strongly attractive two-species fermions trapped in a square optical lattice.  We show that the  strong
interaction and orbital effects can give rise to the emergence of non-trivial lattice structures and pseudo-spin degree of
freedom
by self-assembly of an ultracold gas. The emerged lattice is characterized by topologically protected band crossing points. This effect is counter-intuitive as strong-attraction in general destroys topological order.
As an example of topological properties, we discuss appearance of interaction-driven topological insulating states: Quantum Anomalous Hall
(\textit{QAH}) state characterized by the spontaneously broken time-reversal symmetry with a gap in the bulk
and quantized Hall conductivity and Quantum Spin Hall (\textit{QSH}) state that can be identified as two copies of \textit{QAH} states which on the whole preserves the time-reversal symmetry.

To the best of our knowledge, this is the first proposal to show that non-trivial topological properties are induced by the interplay  between strong interactions and orbital effects. One feature of the present proposal is that our system can contain self-generated 
impurities, domain structures etc due to the spontaneous nature of our emergent lattice. Presence of such imperfections is crucial to observe phenomena such as edge currents or Hall plateaus. This is in contrast to optically created frustrated lattices, where one have to impose additional non-trivial potential to create such imperfections. }

\section{Model}

We consider a mixture of two-species ultracold fermionic atoms trapped in an optical lattice potential
$V_{\sigma, \mathrm{latt}}= V_{\sigma, x}\sin^2(\pi x/a) + V_{\sigma, y}\sin^2(\pi y/a) + V_{\sigma, z}\sin^2(\pi z/a) $,
where $\sigma=\uparrow,\downarrow$ denotes the species and $V_{\sigma,x(y)(z)}$ are the corresponding lattice depths for $\sigma$-fermions along the $x,y,z$ direction
respectively. The lattice constant $a$ is given by the trapping laser wavelength, $a=\lambda/2$. For the two-dimensional ($2D$) geometry
we choose $V_{0}=V_{\downarrow, x}=V_{\downarrow, y}$, 
$V_{1}=V_{\downarrow, z}=V_{\uparrow, x(y)(z)}$, and $V_1 \gg V_0$, so that the $\downarrow$-fermions can effectively move in the $x-y$ plane with
the $z$ motion frozen. Since the $\uparrow$-fermions move in a
deeper lattice, in the first approximation we can neglect the tunneling of these particles. For simplicity we consider the case in which fermionic masses  are equal $m_{\downarrow}=m_{\uparrow}$, which 
implies equal recoil energies  $E_{\rm R}=E_{\rm R\sigma}=\pi^2\hbar^2/2m_{\sigma}a^2$. In this paper we look into a spin-imbalanced
situation with fillings $n_{\downarrow}=1$ and  $n_{\uparrow}=1/2$. It is worth mentioning that such attractive fermion mixtures are already realized in optical lattices for studying superfluidity \cite{Kett}, anomalous transport\cite{EssK, Dem} etc.

Atoms of different types interact with each other via $s$-wave scattering with strength $a_s$. If 
the interaction is strongly attractive ($a_s\ll0$), the $\uparrow$ and $\downarrow$-fermions tend to pair and form composites with
creation operator $\hat{b}^{\dagger}_{\mathbf{i}}=\hat{s}^{\dagger}_{\uparrow \mathbf{i}}\hat{s}^{\dagger}_{\downarrow \mathbf{i}}$ and corresponding number operator $\hat{n}^B_{\mathbf{i}}$ \cite{Micnas, Kuk, Bar}. Here $\hat{s}^{\dagger}_{\sigma \mathbf{i}}, \hat{s}_{\sigma \mathbf{i}}$ are the creation and annihilation operators
of the $\sigma$ fermions in the respective $s$-bands. The composite density is the same as the $\uparrow$-fermions density, i.e. in our case $n_{\uparrow}=n^B=1/2$. 
Such composites are considered static due to the smallness of the tunneling of the minority component. The excess $\downarrow$-fermions 
with filling $m=n_{\downarrow}-n_{\uparrow}=1/2$ can tunnel from one site to another. Recently it has been noted that in the strong interaction regime, the standard Hubbard models
have to be modified due to both intra- and inter-band effects \cite{Om1, Om2, Om3, Dirk}.
Taking these effects into account we construct a minimal model for the composites and the excess $\downarrow$-fermions by including the occupation of the $s$ and $p$-bands and 
the renormalization of the interactions.

\section{Modified Hamiltonian}

The inclusion of the $p$-bands allows one $\downarrow$-fermion to 
occupy the same site as a composite. The single-particle tunneling Hamiltonian then reads: 
\begin{equation}
H_{\rm{t}}= -J_{0}\sum_{\langle\mathbf{ij}\rangle}{\hat s_{\mathbf{i}}}^{\dagger}
\hat{s}^{}_{\mathbf{j}} + J_{1}\sum_{\delta}\sum_{\langle\mathbf{ij}\rangle_{\delta}}{\hat p_{\delta\mathbf{i}}}^{\dagger}\hat p^{}_{\delta\mathbf{j}},
\end{equation}
where $\delta=x,y$ and
$\hat{s}^{\dagger}_{\mathbf{i}}, \hat{s}_{\mathbf{i}} $, $\hat{p}^{\dagger}_{\delta\mathbf{i}}, \hat{p}^{}_{\delta\mathbf{i}}$ are the creation and 
annihilation operators of the $\downarrow$-fermions in the $s$- and $p$-bands respectively. $J_0, J_1 > 0$ are the single-particle
tunneling amplitudes in the $s$- and $p$-band respectively, and $\langle\mathbf{ij}\rangle_{x(y)}$ denotes the nearest-neighbour sites along the $x(y)$-direction. 

The on-site Hamiltonian for the excess $\downarrow$-fermions and the composites including the $s$-and $p$-bands is given by:
\begin{eqnarray}\label{interac}
H_{\rm int} &=& -|U_2|\sum_\mathbf{i} \hat{n}^B_i(1-\hat{n}^{}_{x\mathbf{i}})(1-\hat{n}^{}_{y\mathbf{i}})  \nonumber\\
&-& |U_3| \sum_{\mathbf{i}} \hat{n}^B_\mathbf{i}
(\hat{n}^{}_{x\mathbf{i}}+\hat{n}^{}_{y\mathbf{i}})
- |\delta U_3| \sum_{\mathbf{i}} \hat{n}^{}_{x\mathbf{i}}\hat{n}^{}_{y\mathbf{i}}\hat{n}^B_\mathbf{i}  \nonumber\\
&+& E_1\sum_{\mathbf{i}} (\hat{n}^{}_{x\mathbf{i}}+\hat{n}^{}_{y\mathbf{i}}), 
\end{eqnarray}
where $\hat{n}^{}_{x(y)\mathbf{i}}={\hat p_{x(y)\mathbf{i}}}^{\dagger}{\hat p^{}_{x(y)\mathbf{i}}}$. The renormalized self-energy of the composite is given
by $U_2$ whereas $U_3$ is the strength of the renormalized onsite interactions between a composite and a $\downarrow$-fermion in 
the $p_x$ ($p_y$)-orbital at a given site. $\delta U_3$ denotes the effective three-body interaction between one composite and two $\downarrow$-fermions each in the $p_x$ and $p_y$ orbitals. The origin of the effective three-body interaction comes from the excitations to higher bands. Such higher-body interactions like $\delta U_3$ are already probed in ultracold atom experiments \cite{Bloch1} and are different from the few-body phenomena like three-body bound states arising in Efimov physics \cite{Efimov}.  We find that $\delta U_3$ is small compared to other parameters, so we neglect it at first.
Then the energy cost for an excess $\downarrow$-fermion to occupy the $p$-band of a composite occupied site is given by 
\begin{equation}
\Delta=E_1+(U_3-U_2),
\end{equation}
and can be reduced as one increases the attractive
scattering length. When $\Delta$ is small or negative, the $\downarrow$-fermions can occupy the $p$-orbital of a site with a composite.
\begin{figure}
\includegraphics[width=55mm,natwidth=610,natheight=642]{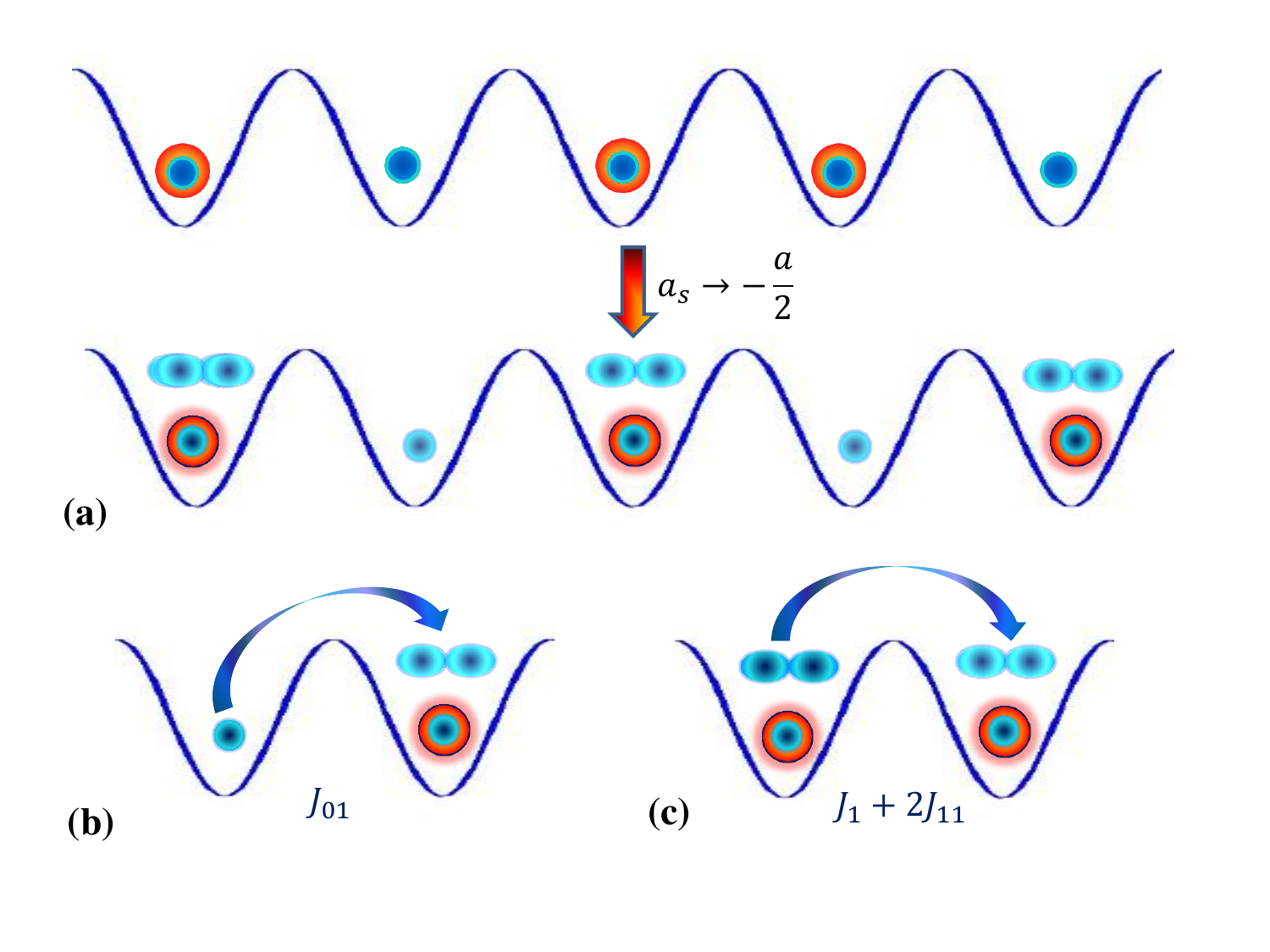}
\caption{\label{fig0} A one-dimensional schematic representation of the system considered in the paper. A red larger sphere refers to a $\uparrow$-fermion and a smaller blue sphere
represents a $\downarrow$-fermion. (a) Top figure: One creates a band insulator for the $\downarrow$-fermions and half-filled system for the $\uparrow$-fermions in presence of weak 
interactions. Bottom figure: Increasing the attractive scattering length $a_s$ leads to an emergence of composites that form a checkerboard structure and the 
remaining $\downarrow$-fermions move between $s$- and $p$-orbitals (semi-transparent blue spheres and dumbells). (b) The interaction induced $s-p$ band hybridization
tunneling element corresponding to Hamiltonian \eqref{sptun}. (c) The effective tunneling in the $p$-band when two composites occupy neighbouring sites.}
\end{figure}

Next we consider two modifications originating from the nearest-neighbour scattering due to the interaction between the excess
$\downarrow$-fermions and the composites. The first of the modifications mixes the $s$- and $p$-bands and can be written as
\begin{equation}\label{sptun}
H_{01} = J_{01} \sum_{\delta=x,y}\sum_{\langle\mathbf{ij}\rangle_{\delta}}\zeta_{i_{\delta},j_{\delta}}{\hat p_{\delta\mathbf{i}}}^{\dagger}\hat{n}^B_\mathbf{i} \hat s_{\mathbf{j}}
+ h.c,
\end{equation}
where $J_{01}$ denotes the interaction induced inter-band tunneling and $\zeta_{i_\delta,j_\delta}=(-1)^{i_\delta-j_\delta}$ reflects the staggered nature 
of the $s-p$ tunneling matrix. This process is shown pictorially in Fig.\ref{fig0}(b) where a $\downarrow$-fermion in the $s$-orbital
is scattered to the $p$-orbital of the neigbouring site due to the interaction with a composite. Such natural non-local hybridization 
between $s-p$ bands due to interaction induced tunneling is an important feature of the strongly interacting gases in lattices. It is
worth to stress that such processes are usually neglected in the literature. Another feature of such $s-p$ hybridization is that due
to parity, any tunneling processes like ${\hat p_{y\mathbf{i}}}^{\dagger}\hat{n}^B_\mathbf{i} \hat s_{\mathbf{j}}$ vanishes for 
$\mathbf{i}=(i_x,i_y)$ and $\mathbf{j}=(i_x\pm 1,i_y)$.

The second modification describes the interaction-induced tunneling in the $p$-band, expressed as:
\begin{equation}\label{inttun}
H_{\rm T} = J_{11}\sum_{\delta}\sum_{\langle\mathbf{ij}\rangle_{\delta}}{\hat p_{\delta\mathbf{i}}}^{\dagger}(\hat{n}^B_\mathbf{i}
+\hat{n}^B_\mathbf{j})\hat{p}_{{\delta\mathbf{j}}}^{},
\end{equation}
where $J_{11}$ denotes intra-band interaction-induced
tunneling for $p_x$($p_y$)-fermions along $x$($y$) directions. $H_T$ gives the most important contribution to the renormalization
of intra-band tunneling \cite{Dirk}. Tunneling in $p$-band is possible only when two neighbouring sites are occupied by composites
(see Fig.\ref{fig0}(c)) as this process conserves energy.
For $a_s<0$, the interaction-induced tunneling $J_{11}$ is negative and the effective tunneling in the $p$-band (given by $J_1+2J_{11}$)
decreases with increasing attraction. Thus in the region where $|J_{01}| \sim |J_1+2J_{11}|$, the excess $\downarrow$-fermions prefer a configuration  with alternating sites occupied by composites {(Fig.\ref{fig0}(b))}. 
The relevant tunneling parameters and interaction parameters are controlled only by the effective interaction 
$\alpha=a_s/a$ and the lattice depths. Their derivation using Wannier functions is discussed in the appendices \textbf{\ref{tunap}}, \textbf{\ref{intap}}.
The magnitudes of the various tunneling amplitudes and $\Delta$ are shown in Fig.~\ref{fig2n}

It is worth to note here that the total Hamiltonian has similar features to the  Falicov-Kimball (FK) model.
Falicov-Kimball model describes interaction between localized classical modes and itinerant quantum modes of a system. 
It was first proposed to study metal-insulator transitions in mixed valence compounds of rare earth and transition metal
oxides \cite{Falicov} and to study crystallization \cite{Lieb}. However there exists one important distinction between the
FK model and our present study. Namely FK models do not possess the correlated multi-orbital tunneling processes. In 
the system that we investigate these processes are not only present but also play a crucial role.

\section{Dynamical Lieb lattice}
In this section, we discus the possible ground state structures of our system characterized by the total Hamiltonian $H=H_{t}+H_{\rm T}+H_{ 01}+H_{\rm int}$. The total Hamiltonian $H$ does not contain composite tunneling and  the commutator $[\hat{n}^B_{\mathbf{i}},H]=0$. Therefore $n^B_{\mathbf{i}}=0,1$ becomes a good quantum number.
We find the ground state of the system by comparing the energies of different configurations of $n^B_{\mathbf{i}}$
over the entire lattice. The search space is too large to compare the energy of every single configuration. Therefore we 
locate a good approximation to the global optimum by using simulated annealing
\cite{Kirkpatrick,AnnealingBook} (For details see appendix \textbf{\ref{simu}}).
We find the lowest energy configurations of the composites for various parameters
on a $12\times 12$ lattice with periodic boundary conditions. While calculating the energy for every single configuration,
we take into account weak attraction between the orbitals (Eq.\eqref{interac}) using Hartree-Fock approximation.
The resulting phase-diagram for the composites is shown in Fig.\ref{fig1}(a). We distinguish the following phases for the composites:
\begin{figure}
\begin{center}
\vspace{-0.25cm}
\includegraphics[width=80mm,natwidth=610,natheight=642]{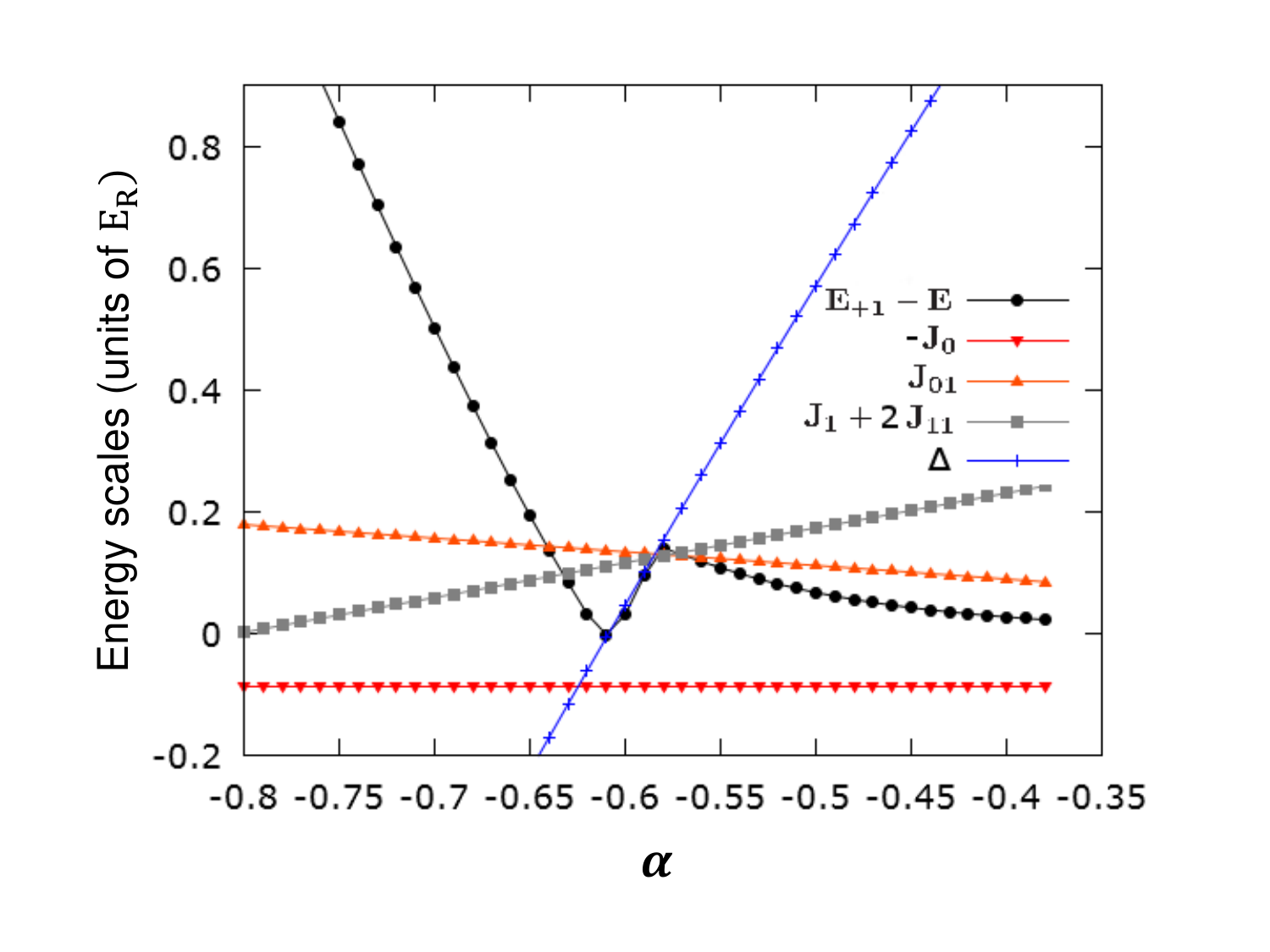}
\caption{\label{fig2n} Comparison of different energy scales present in the system as a function of the effective interaction strength. 
Color lines represent different hopping parameters. Black line with dots 
represent the energy difference between state of a system with half filling of $\uparrow$-fermion ($\mathbf{E}$) and a state with
one $\uparrow$-fermion more than half filling ($\mathbf{E_{+1}}$). Energies are presented in the unit of $E_R$ and lattice depths are 
$V_0=4E_{\rm R}$, $V_1=20E_{\rm R}$}
\end{center}
\end{figure}
\begin{figure*}
\begin{center}
\includegraphics[width=160mm,natwidth=610,natheight=642]{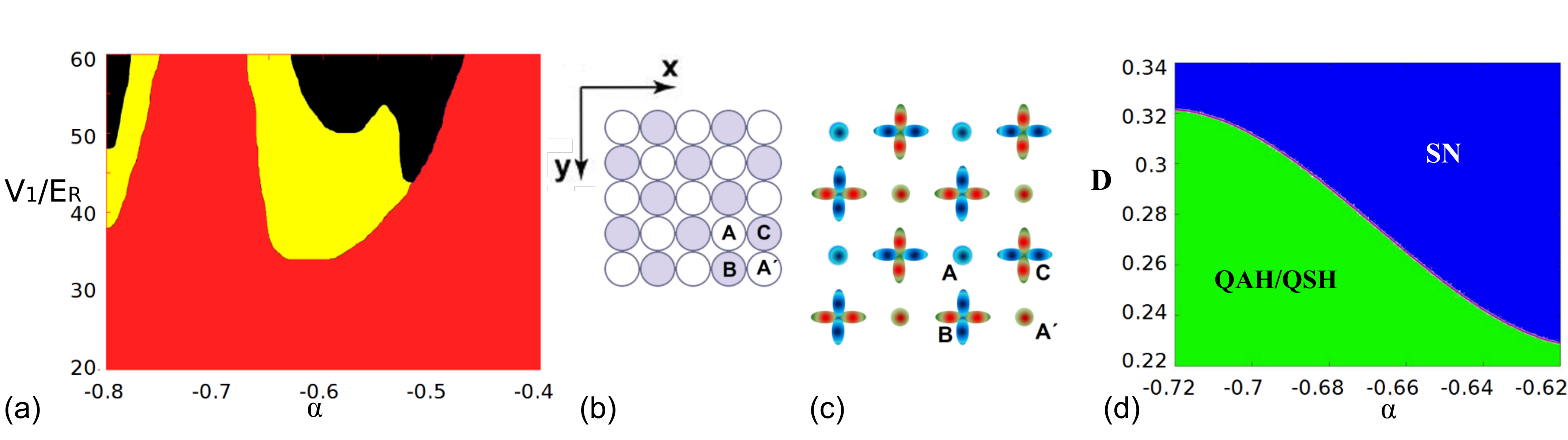}
\caption{\label{fig1} (a) The phase diagram for the configurations of the composites. The red coloured region denotes the period-1 checkerboard (CH1) configuration for the composites. The black 
region denotes the phase-separated state, the yellow region contains the mixed phase. The shallow lattice parameter
is $V_0=4E_{\rm R}$. (b) Distribution of the composites in the \textit{CH1} phase. The filled (empty) circles denote the presence (absence) of a composite.
(c) Orbital degree of freedom for the excess $\downarrow$-fermions in CH1 lattice. The coloured circles denote $s$-orbitals. The horizontal (vertical) coloured
dumb-dell shapes denote $p_x (p_y)$-orbitals. The basis states for the blue (red) lattice are denoted by $\mathbf{A}, \mathbf{B}$ and  $\mathbf{C}$
($\mathbf{A'}, \mathbf{B}$ and  $\mathbf{C}$). The corresponding sites in the \textit{CH1} structure are shown in Fig.~2(b). (d) The ground state phase diagram for the excess $\downarrow$-fermions corresponding to 
the composite \textit{CH1} phase. The phases are shown as a function of dipolar strength $D$ and
contact interaction strength $\alpha$ for $V_0=4E_{\rm R}$. The blue region denotes spin-nematic (SN) phase whereas the green region denotes Quantum Anomalous Hall/Quantum Spin Hall (QAH/QSH) phases.}
\end{center}
\end{figure*}
i) checkerboard structure with period one (\textit{CH1}; Fig.\ref{fig1}b)) 
ii) mixed phase characterized by the absence of any periodic structure and iii) the phase-separated
state characterized by the clustering of the composites to one region of the lattice. The mixed phase occurs in the region where 
the energy cost to occupy the $p$-orbital is small compared to other tunneling processes. Thus it is possible that the mixed phases 
contain self-generated disorder due to the composite density dependence on the tunneling processes. 
We have checked the existence of the mixed phases and obtained phase-boundaries also for lattice size of $8\times 8$ and $16 \times 16$ and the phase boundaries remain qualitatively unchanged. 

The {\it CH1} region is the most interesting one with respect to generation of non-trivial topological lattices.
{In the parameter regime, for lattice depth $V_1 \lesssim 35E_R$ with $V_0=4E_R$ we find that,
the \textit{CH1} region becomes the ground state. In the rest of our paper we will then concentrate on this particular parameter space.}
The presence of \textit{CH1} region can be qualitatively predicted for $\Delta\ll 0$ {and} it can be easily shown that the 
\textit{CH1} structure has the lowest energy provided that $2J^2_{01}/|\Delta| > |J_{1}+2J_{11}|/\pi$. {On the other hand, for 
$\Delta > 0$, \textit{CH1} structure has the lowest energy as long as $ J^2_{01}/|\Delta| \ne 0$.}

{We also note that origin of \textit{CH1} structure is different from the origin of the anti-ferromagnetic Neel phase for 
the repulsive Fermi-Hubbard model. For the repulsive fermions, the Neel state  
arises in the balanced mixture due to the lowering of energy in the form of second order exchange processes due to tunneling induced localized creation of pairs. On the other hand, in the present case, the excess fermions are delocalized over the whole lattice. Then the {\it CH1} structure appears as a result of the minimization of the total kinetic energy of the delocalized excess $\downarrow$-fermions.} Moreover we 
examine the energy cost related to the addition of the minority component. In Fig.~\ref{fig2n}, we plot the energy cost to dope the 
\textit{CH1} phase with an additional minority component for lattice depths $V_0=4E_{\rm R}$, $V_1=20E_{\rm R}$. The energy cost is denoted by $\mathbf{E_{+1}}-\mathbf{E}$. We see that in the regime of
$\Delta<0$ it costs additional energy of the order $\sim |\Delta|$ to dope with a minority component. In the regime of negative
$\Delta $, this energy cost, $\mathbf{E_{+1}}-\mathbf{E}$, is much larger than the other tunneling processes (Fig.~\ref{fig2n}). The \textit{CH1} phase is then robust against small doping of minority components. 

Now, we focus our attention on the behaviour of the excess mobile fermions. The excess fermions move on the {\it CH1} 
structures created by the composites. Considering the distribution of the orbitals that excess fermions can occupy, the motion of these particles can be divided into two sub-lattices presented by blue and red colour in Fig.\ref{fig1}(c). In order to see this,  let us consider an empty site $A$ shown at the composite structure in Fig.\ref{fig1}(b). The $s$-orbital of this site (shown as the blue site denoted by $A$ in Fig.\ref{fig1}(c)) can be occupied by an excess fermion. Then the fermion occupying the site A can either move to the $p_x$-orbital of the $B$ site or the $p_y$-orbital of the $C$ site under the influence of the Hamiltonian \eqref{sptun}. This is due to the fact that both $B$ and $C$ sites are occupied by composites as denoted by dark circles in Fig.\ref{fig1}(a). Then due to the absence of any tunneling matrix element between $p_x$-orbital of site $B$ ($p_y$-orbital of the $C$) to 
$s$-orbital at site $A'$, the excess particles will only move in the blue sub-lattice as shown in Fig.\ref{fig1}(c). Similarly one can construct the red sub-lattice geometry. This takes place because of the directional nature of the inter-orbital tunneling $J_{01}$ in the Hamiltonian \eqref{sptun} and the absence of any on-site orbital mixing term in \eqref{interac} due to {parity and fermionic statistics.} Each of the sub-lattices in Fig.\ref{fig1}(c) can be characterized by three basis sites denoted by $\mathbf{A}, \mathbf{B}$ and  $\mathbf{C}$ (for the blue lattice) and
$\mathbf{A'}, \mathbf{B}$ and  $\mathbf{C}$ (for the red lattice). Both, the red and the blue sub-lattices have the structure of a Lieb lattice 
 \cite{Lieb1}.
Let us denote the excess $\downarrow$-fermions moving in the blue sub-lattice by
${\Phi_1}=[\hat{s}_A, \hat{p}_{\rm y B}, \hat{p}_{\rm x C} ]$ and in the red sub-lattice 
by $\Phi_2=[\hat{s}_{A'}, \hat{p}_{\rm x B}, \hat{p}_{\rm y C}]$.  
We can see that, due to the interaction, we 
 induce one pseudo-spin degree of 
freedom in the form of orbitals in different sub-lattices. Their motion is governed by the Hamiltonian:
\begin{eqnarray}\label{Hamlieb}
H &=& J_{01} \left [\sum_{\langle\mathbf{ij}\rangle_x}\zeta_{i_x,j_x}\hat{s}^{\dagger}_{\rm A \mathbf{i}} {\hat p_{\rm xC\mathbf{j}}}^{}
+\sum_{\langle\mathbf{ij}\rangle_y}\zeta_{i_y,j_y}\hat{s}^{\dagger}_{\rm A \mathbf{i}} {\hat p_{\rm yB\mathbf{j}}^{}} \right . \nonumber\\
&+& \left . \sum_{\langle\mathbf{ij}\rangle_x}\zeta_{i_x,j_x}\hat{s}^{\dagger}_{\rm A' \mathbf{i}} {\hat p_{\rm xB\mathbf{j}}^{}}
+ \sum_{\langle\mathbf{ij}\rangle_y}\zeta_{i_y,j_y}\hat{s}^{\dagger}_{\rm A' \mathbf{i}} {\hat p_{\rm yC\mathbf{j}}^{}} + h.c \right] \nonumber\\
&+& \Delta \sum_\mathbf{i,\tau=B,C} (\hat{n}_{\rm \tau x\mathbf{i}}+\hat{n}_{\rm \tau y\mathbf{i}}) -
|\delta U_3| \sum_{\mathbf{i},\tau=B,C} \hat{n}_{x\tau\mathbf{i}}\hat{n}_{y\tau\mathbf{i}}. \nonumber\\
&&
\end{eqnarray}
Here the first term (the one inside the [.]-bracket) in Eq.\eqref{Hamlieb} is a reformulation of $H_{01}$ {from} Eq\eqref{sptun}. The second term refers to the energy cost of the $p$-orbital
atoms occupying a site already taken by a composite. The third term describes effective onsite interactions between the red and blue
fermions on the sites $B$ and $C$. 
\begin{figure}
\begin{center}
\hspace{-2.0cm}
\includegraphics[width=105mm,natwidth=610,natheight=642]{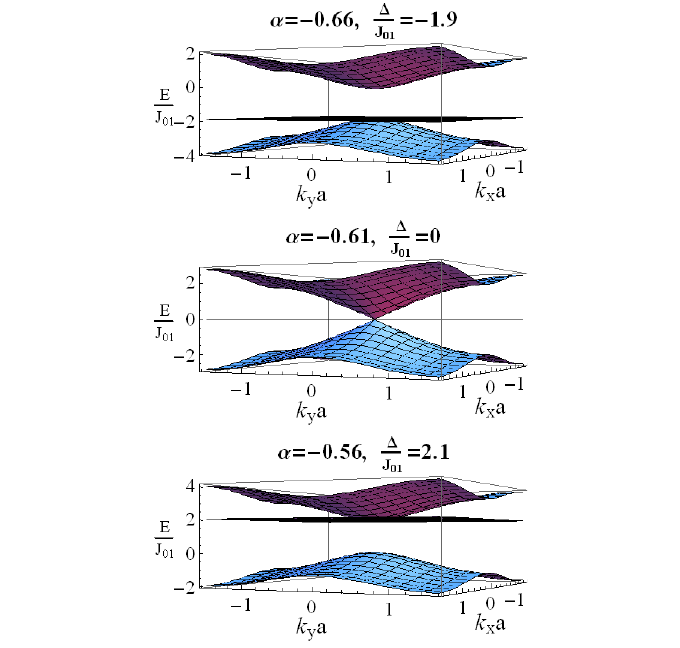}
\caption{\label{disp} We show the dispersion relations of the Lieb lattice, as expressed in \eqref{disp}, for three different values of $\Delta$ with lattice depths $V_0=4E_{\rm R}$, $V_1=20E_{\rm R}$}.
\end{center}
\end{figure}
Now, let us focus on the single particle dispersion relation. Diagonalizing the single particle part of Hamiltonian \eqref{Hamlieb}, 
we get
\begin{equation}\label{disp}
\epsilon_{\mathbf{k}}\in\{\Delta, \Delta/2 \pm \sqrt{\left(\Delta/2\right)^2+4J^2_{01}\left[\sin^2 k_xa + \sin^2 k_ya \right]}\},
\end{equation}
where the momentum $\mathbf{k}=(k_xa,k_ya)$ belongs to the reduced Brillouin zone $(-\pi/2,\pi/2)$. The dispersion relation in 
Eq.\eqref{disp} is plotted in the Fig.\ref{disp} for three different $\Delta$. The dispersion contains a quadratic band crossing point (QBCP) for $\Delta \ne 0$ with one of the dispersive bands touching the flat band at momentum $(0,0)$. For $\Delta=0$, three bands touch each other at the momentum $(0,0)$ with the upper and lower band having linear dispersion in the vicinity of this point.
For simplicity let us consider only the case when $\Delta < 0$. {We can write an effective two band Hamiltonian $H=d_0\mathcal{I}+d_z\sigma_z+d_x\sigma_x$, where $\sigma_{x(z)}$ are  the Pauli matrices, $\mathcal{I}$ is the identity matrix,
$d_0=-(J^2_{01}/\Delta) (\cos 2k_xa + \cos 2k_ya)$ and the vector $\vec{d}=(d_x,d_z)=-(J^2_{01}/\Delta) (4\sin k_xa \sin k_ya, \cos 2k_xa - \cos 2k_ya)$. In this limit,
the particles occupy only the $B$ and $C$ sites of the lattice and the  population in the $A$ and $A'$ sites is negligible. {For excess fermion filling $m=1/2$, the dispersive band is filled and any excitation to the next flat band remains localized in space. This makes the system an insulator.} The next dispersive band is separated by an energy gap of $\Delta$ at $\mathbf{k}=(0,0)$.
We introduce a normalized vector  $\hat{d}=\vec{d}/\sqrt{|\vec{d}|}$ (mapping from the Brillouin zone
to a 2-sphere) and define the Berry phase as:
\begin{equation}
B=\int d^2\mathbf{k}\  \hat{d} \cdot \left[\partial_{k_x}\hat{d} \mathbf{\times} \partial_{k_y} \hat{d} \right ],
\end{equation}
with the integration over the Brillouin zone. The vector $\vec{d}$ acts like an effective magnetic field and the
corresponding Berry phase is given by $\pm 2\pi$.} Appearance of non-zero Berry phase makes the QBCP topologically stable. Moreover, 
at excess fermion filling $m=1/2$, the lower band is completely filled. The system remains insulating with the topological lower
band filled. In this regard, the Lieb lattice is different from the honeycomb lattices where near the band crossing
points (Dirac points), one has linear dispersion relations. Because of the presence of 
dispersive band at the crossing point, unlike Dirac point, the system is unstable towards topologically insulating states
even for the introduction of very small spin-orbit coupling.

We would like to point out that such emergent non-trivial lattice structures are not possible even if one realizes \textit{CH1}-structures
in different systems such as dipolar systems or atomic mixtures \cite{Lew2}. Even in the Falicov-Kimball model, due to the absence
of such inter-orbital hybridization, a {\textit CH1} structure can not give rise to a non-trivial lattice geometry for the mobile
fermions. Summarizing, in this section we have shown that, due to interaction-induced tunneling, one can 
generate topologically protected exotic lattices starting from trivial geometries. 

\section{Proposed experimental realization}

For experimental realization of the present proposal, we consider a band insulator for $\downarrow$-fermions and half-filling
for $\uparrow$-fermions trapped on a square lattice where the inter-species interaction is weak. Such species dependent 
lattices were already experimentally realized to study glassy behaviour, as in ref.\cite{Schneble}. Then by increasing the scattering
length in the attractive regime via Feshbach resonance (or confinement-induced resonance), one can reach the regime of
a dynamical Lieb lattice. As one gets to the region with $\Delta \approx 0$, the Lieb lattice emerges due to
the {\it CH1} structure of the composites. {To experimentally detect this phase, one can probe the excitation 
spectrum of the mobile fermions using Bragg spectroscopy \cite{stenger, davidson} or by using momentum-resolved intra-band transitions 
\cite{tar}. Such measurements can show signature of the Lieb-lattice structure by showing the the presence of QBCP and the curvature of various bands. Additionally, measurement of the density-density correlations from the expansion of the minority
component can give a signature for the {\it CH1} structure \cite{luk} arrangement of the composites.} Due to the appearance of the \textit{CH1} over a wide range of lattice depths and scattering lengths, as shown in Fig.\ref{fig1}a, it is indicative that this result is stable under small changes of parameters.

{Next, we briefly discuss the role of tunneling of the minority $\uparrow$-fermions for experimental realization. 
The effective tunneling strength (denoted by $J_{\uparrow}$) of the $\uparrow$-fermions includes both single-particle tunneling as well
as contributions from the interaction. To reach the {\it CH1} configuration, the tunneling of the $\uparrow$-fermions is important
as it helps to scan the large set of possible configurations for the composites. In the present situation, 
(as depicted in Fig.\ref{fig0}(a)) for relatively weak interaction ($\Delta>0$), due to the high density imbalance, almost
every composite has an excess $\downarrow$-fermion as a nearest neighbour. Therefore, due to tunneling of the minority component,
a composite can effectively move to a neighbouring site that already contains an excess $\downarrow$-fermions. Thus the time scale
required to reach the {\it CH1} lattice configuration is set by the minority component tunneling rate. {Moreover, to generate the long-range order over the entire system, one needs many such tunneling events. Subsequently the timescale to form the entire {\it CH1} configuration will be set by the corresponding Lieb-Robinson bound. In that situation, within the timescale allowed by the loss rates, domains of {\it CH1} order with different orientation will be created}. 
For the lattice depths $V_0=4E_{\rm R}$, $V_1=20E_{\rm R}$ and interaction strength $a_s/a=-0.5$, we have found that the timescale for the {\it CH1} pattern to occur is of the order of $\sim 1$ms. It is worth nothing that the situation here is different from the spin-balanced case. In the spin-balanced situation, the composites can only have a vacant neighbour where they can hop 
via slow second-order process with strength $\sim J_0J_{\uparrow}/|U_2|$ resulting in slower redistribution of the pairs \cite{EssK, Hofs}. 
Due to the presence of the hopping of the $\uparrow$-fermions, our calculations in the previous section is valid as long as 
$J_{\uparrow} \ll \{|J_{01}|,J_0,J_{1}+2J_{11}\}$. We have calculated that the various tunneling terms of the excess fermions (specially
$J_{01}$) are at least one order of magnitude larger than $J_{\uparrow}$. Because of the separation of tunneling scales between the excess fermions and the composites, one can use Born-Oppenheimer like approximation and recover the FK-like Hamiltonian discussed in the present and previous sections.}

Regarding the relevant atomic species for such experiments, one such choice 
could be fermionic ${}^6$Li species or fermionic ${}^{40}$K. For a lattice constant of $a=500$nm, 
the corresponding scattering length is on the order of $a_s \sim -300$nm. This is already achieved in Lithium mixtures in 
Refs.\cite{Salomon, Grimm, ketterle} and fermionic Potassium mixtures in Refs. \cite{Jin}. The other option is a 
mass-imbalanced mixture. In such case the effective scattering length is scaled and $\alpha \approx (a_s/a)(1+m_{\downarrow}/2m_{\uparrow})$ for the same parameters
as used in the case of equal mass.  Thus, if one traps ${}^{40}$K in the weaker lattice of $V_0=4E_{\rm R\downarrow} $ ($\downarrow$-fermions) and ${}^6$Li in the stiffer lattice of $V_1=20E_{\rm R\uparrow}$, then the Lieb lattice phase can be obtained for a scattering length of $a_s$ (KLi) $\approx -80$nm. Such a strongly attractive
scattering length can be experimentally realized using the narrow Feshbach resonance for ${}^{40}$K-${}^6$Li mixture by tuning
the magnetic field at the milli-Gauss accuracy \cite{pietro}. {The possible temperature range to achieve Lieb lattice structure is determined by the bandwidth of the excess mobile fermions. For a scattering length of $a_s/a \sim -0.5$, Lieb lattice phase is achievable 
as long as the temperature is lower than $J_{01} \sim 0.1E_R$. For Lithium mixtures this translates to a temperature scale of $\sim 
100$nK and for Potassium-Lithium mixtures the corresponding temperature is $\sim 20$nK.} 

{One important process that can hinder experimental realization of the present scheme is the heating due to photon scattering in a deep optical lattice for the $\uparrow$-fermions. It is known that the optical lattice depth is proportional to $\sim (\delta\omega)^{-1} $ and photon scattering rate is proportional to $ \sim (\delta\omega)^{-2}$ where $\delta\omega$ is the detuning of the laser frequency. For a far-detuned laser creating the shallow optical lattice for the $\downarrow$-fermions, from Ref.\cite{DeMarco} we took the heating rate, $\dot{T}_{\downarrow}\approx10^{-4}E_R/$ms for ${}^6$Li and $\approx 5.10^{-5}E_R/$ms for ${}^{40}$K for laser wavelength of $1064$nm. 
Then using the relation between the lattice depth, photon scattering rate and detuning, one can find an 
estimate for the heating rate in the deeper lattices from the ratio, $\dot{T}_{\uparrow}/\dot{T}_{\downarrow}\approx (V_1/V_0)^2$. For 
lattice depths of $V_1=20E_R$ and $V_0=4E_R$ we find that $\dot{T}_{\uparrow} \sim 0.002 E_R/$ms for ${}^6$Li and 
$\approx 0.001E_R/$ms for ${}^{40}$K. As the bandwidth of the Lieb lattice is on the order of $\sim 0.1E_R$, this restricts the duration of the experiments to $\sim 100$milliseconds for both Lithium and Potassium mixtures.}
The limiting effect of radiative losses, in principle, could be eliminated by using alkaline-earth atoms, 
like Ytterbium. Ytterbium does not allow for magnetic Feshbach resonances, but can permit confinement induced resonances
in ultra tight traps \cite{bruno}. Yet another, so far relatively unexplored option could be to use
alkali-earth alkali mixtures like Ytterbium-Lithium \cite{subha}. Recently it has been proposed that 
due to hyperfine coupling between the electron spin and nuclear magnetic moment, magnetic Feshbach resonance (with width $\sim 2.8$mG) will occur between the ground state fermionic Ytterbium and Lithium atoms \cite{Hutson}. 

{ Next, we examine the effect of two-body and  three-body inelastic loss processes. 
Due to the anti-commutation relation between the fermions of the same species (irrespective of the orbitals they occupy), 
the three-body loss from the s-wave collisions vanishes. Then the two-body collisions become the dominant loss process.  To look into a particular example, we choose ${}^{40}$K$-{}^{6}$Li mixture, where two-body losses occur due to spin-relaxation \cite{Grimm2}. We define the two-body 
decay rate as $L=L_2 \int {\cal W}^s_{\mathbf{i},\downarrow}(\vec{r}) {\cal W}^s_{\mathbf{i},\uparrow}(\vec{r})dr$ where
${\cal W}^M_{\mathbf{i},\sigma}(\vec{r})$ are Wannier functions on site $i$, for a species $\sigma$ on a band $M$ and  
$L_2$ denotes the two-body loss rate. Then the particle loss rate is given by $N(t)=N(0)\exp\left[-L t\right]$, and the corresponding 
lifetime is $\sim L^{-1} \sim 1$s for $L_2 \sim 10^{-13}cm^3/s$ and lattice constant of $a \sim 500$nm. Also for Feshbach resonances in the  ground state alkali-earth alkali mixtures \cite{Hutson}, such two-body loss processes will be absent.}

{We conclude this section by discussing briefly the effect of impurities which can appear in experimental realizations of 
\textit{CH1} structures. Such impurities can appear in the form of excess composites or missing composites in the \textit{CH1} structures.
These defects are reminiscent of the interstitial defects in solid-state crystals. Presence of such defects will create 
local regions with tunneling between $p-p$ orbitals with strength $\sim J_p=J_{1}+2J_{11}$ or tunneling between the $s-s$ orbitals
with strength $\sim J_0$. In the limit of dilute impurities, one can estimate the effective impurity strength as
$g_{\rm imp}\sim n_{\rm imp} {\rm max}[J^2_p, J^2_0]/W^2$, where $n_{\rm imp}$ is the impurity density with $n_{\rm imp} \ll 1/2$ and
$W$ is the bandwidth of the clean lattice. From Eq.\eqref{disp} we see that when $\Delta \sim 0$, the bandwidth is given by $ W\sim J_{01}$ and when $\Delta \ll 0$, the bandwidth changes to 
$W\sim J^2_{01}/\Delta$. Now as long as $g_{\rm imp} \ll 1$, one can recover the clean limit of the dispersion relation with 
the density of states for the flat band showing a width on the order of $g_{\rm imp}$ \cite{vigh}. Assuming a impurity 
concentration of $n_{\rm imp}=0.05$, we find that $g_{\rm imp}\sim n_{\rm imp} =0.05$ for $\Delta=0$ and $g_{\rm imp}\sim 2n_{\rm imp} =0.1$ for $\Delta\approx 2J_{01}$ for $a_s/a\sim -0.7$ for the parameters shown in Fig.~\ref{fig2n}. Thus we find that dilute impurities will have negligible effects on the properties of the Lieb lattices.}

\section{Dynamical topological insulators}
The dynamical realization of a Lieb lattice opens up an alternative way to study the possibility of generating
integer quantum Hall effects with cold atom systems.
{One possible way to generate quantum Hall states such as Quantum Anomalous Hall (\textit{QAH}) and Quantum Spin Hall 
(\textit{QSH} states) in nontrivial lattices is by inducing
effective spin-orbit coupling \cite{Science1, Science2}. Such coupling can be achieved through optical means \cite{Dali}, by lattice-shaking \cite{andre} or dynamically by including long-range interactions \cite{Raghu, Fradkin}. In our proposals we use the last of these methods and
the effective spin-orbit coupling is induced by the mean-field effect of the long-range interaction. }

{Models with long range interactions are usually hard to implement in an experimentally realizable system, as
the on-site interaction has to be of the same order of magnitude as the long-range part \cite{Honerkamp}.} To investigate such possibility in our system we add an extra magnetic dipolar term (restricting its range to next-nearest neighbours)
for the excess $\downarrow$-fermions,
\begin{eqnarray}\label{fullham}
H_{\rm full} &=& H + H_{\rm dd}, \nonumber\\
H_{\rm dd} &=& U_{\rm dd} \sum_{\mathbf{i},\tau} \hat{n}_{x\tau\mathbf{i}}\hat{n}_{y\tau\mathbf{i}} +
\frac{U_{\rm xy}}{2}\sum_{\langle\langle\mathbf{i,j}\rangle\rangle,\tau \neq \tau'} \hat{n}_{x\tau\mathbf{i}}\hat{n}_{y\tau'\mathbf{j}} \nonumber\\
&+& \frac{U_{\rm
xx}}{2}\sum_{\langle\langle\mathbf{i,j}\rangle\rangle,\tau }
[\hat{n}_{x\tau\mathbf{i}}\hat{n}_{x\tau\mathbf{j}}+\hat{n}_{y\tau\mathbf{i}}\hat{n}_{y\tau\mathbf{j}}],
\end{eqnarray}
where $U_{\rm dd}$ is an onsite dipolar interaction, $U_{\rm xy}$ is an
interaction between the particles in $p_x$ and $p_y$-orbital in $B$
and $C$ sites respectively, and $U_{\rm xx}$ is a next-nearest
neighbour interaction between the particles in $p_x$ and
$p_x$-orbital (also between $p_y$ and $p_y$ orbital) in $B$ and $C$
sites. $\langle\langle\mathbf{i},\mathbf{j}\rangle\rangle$ denotes
next-nearest neighbour $p$-orbital sites. We additionally introduce
the dimensionless dipolar interaction strength
$D=\mu_0\mu^2m_{\downarrow}/2\hbar^2a$, where $\mu$ is the magnetic
dipole moment of the atoms and $\mu_0$ is the vacuum permeability.
The dipole-dipole interaction has the form $U_{\rm
dd}(r)=D (1-3z^2/r^2)/r^3$, where $r$ is the inter-particle
distance. {Effectively the fermions have a two-dimensional nature so all the dipolar interaction terms are repulsive.}
For experimental realization, the suitable candidates are:
fermionic ${}^{161}$Dy, which is experimentally available in a quantum
degenerate state \cite{Lev}, and fermionic ${}^{167}$Er
\cite{Inns}. {Dy and Er can also be suitable due to the possibility to achieve lattices with laser wavelengths $\sim 400$nm as
discussed in Ref.~\cite{Lev}. This will reduce the s-wave scattering length needed to achieve the emergent Lieb lattice phase to 
$a_s \sim -100$nm. Although - due to the presence of a zoo of Feshbach resonances in these atoms - one probably needs high tunability
of the magnetic field.} {One can also use polar molecules} 
provided that the short range interaction is modified, for instance using confinement-induced resonances.

Due to the strong attractive contact interaction $|U|$, the effect of dipolar terms on $\Delta$ is negligible. Moreover, we
neglect the effective long-range repulsion between the composites
which can further stabilize the dynamical Lieb lattice phase. {Then} within the
weak-coupling limit the mean-field parameters can be defined:
\begin{eqnarray}\label{mf}
&&\langle \hat{p}^{\dagger}_{xBi}\hat{p}^{}_{yCj}\rangle=\langle\hat{p}^{\dagger}_{yBi}\hat{p}^{}_{xCj}\rangle= i\chi_{\rm QAH}^{},  \nonumber\\
&&\langle\hat{p}^{\dagger}_{xBi}\hat{p}^{}_{yCj}\rangle=-\langle\hat{p}^{\dagger}_{yBi}\hat{p}^{}_{xCj}\rangle=i\chi_{\rm QSH}^{},  \nonumber\\
&&\langle\hat{n}_{xBi}\rangle-\langle\hat{n}_{yCj}\rangle=\langle\hat{n}_{xCi}\rangle-\langle\hat{n}_{yBj}\rangle = \chi_{\rm SN}^{}, \nonumber\\
&&
\end{eqnarray}
where $\chi_{\rm QAH}^{}$ denotes the order parameter for the
\textit{QAH} state. QAH is characterized by a loop-current,   broken
time-reversal symmetry (TRS) and topologically
protected chiral-edge states. The \textit{QSH} state order parameter $\chi_{\rm QSH}^{}$ can be thought of as two copies of \textit{QAH} which on the whole conserve time-reversal symmetry \cite{Zhang2}. This state contains helical edge states as shown in \cite{Wu}.
We see that the mean-field effect of the interaction effectively creates a spin-orbit coupling.
The last order parameter $\chi_{\rm
SN}^{}$ refers to the spin-nematic state (SN). It breaks $C_4$ symmetry between the blue and red sub-lattices
and constitutes an anisotropic semi-metal \cite{Fradkin}. {Near the QBCP point, the mean-field energy is then given by
\begin{equation}
E_{mean}=-\sum_{\mathbf{k}}E_{\rm \mathbf{k}} + \left[U_{\rm xy}+\frac{U_{\rm dd}-|\delta U_3|}{4}\right]\chi^2_{\rm SN}+U_{\rm xy}\chi^2,
\end{equation}
where the dispersion relation is given by $E_{\rm \mathbf{k}}=\sqrt{\left[(k^2_y-k^2_x)-\left(U_{\rm xy}+\frac{U_{\rm dd}-|\delta U_3|}{4}\right)\chi_{\rm SN}\right]^2+4k^2_xk^2_y+U^2_{\rm xy}\chi^2}$ and the order parameter $\chi=\chi_{\rm QAH}$ or $\chi=\chi_{\rm QSH}$. 
Then we find various order parameters by minimizing the mean-field energy $E_{\rm mean}$}. We have calculated how these order parameters change with interaction strength $\alpha=a_s/a$ and dipolar strength $D$. The obtained phase diagram is presented in Fig.\ref{fig1}(d). We see that for lower dipolar strength $D$ one can stabilize Quantum Hall states whereas for higher dipolar strength, the spin-nematic state minimizes the energy. This can be qualitatively explained by the fact that for
weaker values of $D$, the repulsive dipolar onsite energy ($U_{\rm dd}$) in Eq.\eqref{fullham} is compensated by the effective onsite 
attraction $\delta U_3$ in Eq.\eqref{Hamlieb}. Consequently, the mean-field physics is dominated by the long-range part of the
dipolar terms which results in stabilization of the \textit{QAH}/\textit{QSH} states. {Even a small dipolar strength
will make the system unstable towards the \textit{QAH}/\textit{QSH} states, but the gap in the bulk will be exponentially small. In that case, one need very low temperature to observe such phases.}
On the other hand, for much higher dipolar strength, the repulsive on-site energy dominates the other interactions, which in-turn stabilizes the spin-nematic phases. Within the mean-field ansatz \eqref{mf} both \textit{QAH} and \textit{QSH} have the same energy, although this
degeneracy can be broken by including higher order exchange
interactions \cite{Raghu}. 
The corresponding mean-field transition temperature to the
\textit{QAH}/\textit{QSH} state is given by, $T_c \sim
(4J^2_{01}/\Delta) \exp[-J^2_{01}/2U_{\rm xy}\Delta] \sim 0.01E_R$
for $D=0.29$ and $\alpha=-0.7$. Such dipolar strength can be reached in fermionic Dysprosium with lattice constant of $a=500$nm
and by fermionic Erbium with lattice constant of $a=300$nm.

\section{Conclusion}

In conclusion, we have presented a theoretical proposal on how frustrated lattices can be created as an effect of self-assembly of cold-atoms. We believe that our proposal opens up another fascinating route for experimental and theoretical studies of frustrated systems. The proposed scheme is very general and can be extended to other lattice structures even in three dimensions. Moreover, by varying the fermionic densities one can get different composite structures where different lattice geometries can be realized by the moving excess $\downarrow$-fermions. On the other hand our proposal gives potential facilitation for the experimental realization of topological insulator. Namely it does not involve additional optical components other than the ones needed for creating the parent lattice.

\begin{acknowledgments}
We would like to thank Jakub Zakrzewski for simulating discussions and Micha\l{} Maik for providing valuable suggestion to improve the 
manuscript. We acknowledge the support by the EU STREP EQuaM, IP AQUTE and SIQS, ERC Grant QUAGATUA, AAII-Hubbard. O. D. also acknowledges support
from National Science Centre Poland project DEC-2012/04/A/ST2/00088. A. P. is supported by the International PhD Project
"Physics of future quantum-based information technologies", grant MPD/2009-3/4 from Foundation for Polish Science and by the University of Gdansk grant
BW 538-5400-0981-12. A.P. also acknowledges hospitality from ICFO. Calculations were carried out at the Academic Computer Center in Gdansk.

\end{acknowledgments}

\appendix

\section{Derivation of $J_{01}$ and $J_{11}$ in the modified Hamiltonian}\label{tunap}
\begin{figure}
\centering
\includegraphics[scale=0.35]{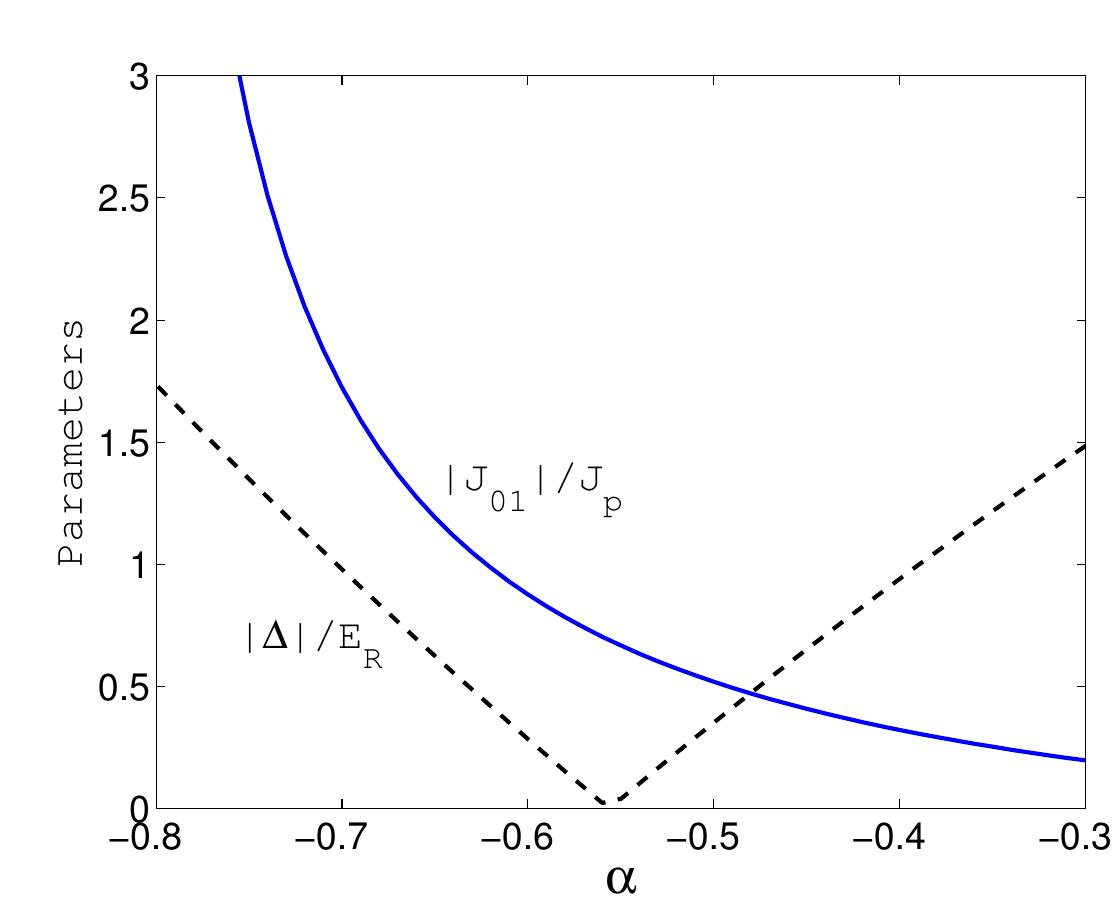}
\caption{ Important parameters in our paper:  the energy cost $|\Delta|/E_R$ (black dashed line) and the relative strength of the $s-p$-band tunneling
and the effective tunneling in $p$-band, $J_{01}/|J_p|$ (the blue solid line), as a function of the effective interaction strength $\alpha=a_s/a$. 
We fix the $\downarrow$-fermion lattice depth $V_0=4E_R$ and the $\uparrow$-fermion lattice depth $V_1=30E_R$ for which we see from Fig.(2) in the paper that the {\it CH1} state
is stable. For low $\alpha$, $\Delta$ is positive. As $\alpha$ becomes more negative, $\Delta$ decreases. For $\alpha \lessapprox -0.56$, $\Delta$ becomes negative and
it`s absolute value increases. From the blue curve we also see that with increase in
$|\alpha|$, the effective $p$-band tunneling decreases and $s-p$ tunneling increases resulting in an increase in the ratio $J_{01}/J_p$. Around $\alpha \sim -0.6$ contribution of
each tunneling processes become of the same order of magnitude facilitating the stability of the {\it CH1} phase in the paper.}
\label{Supfig1} 
\end{figure}

Here we describe the procedure to calculate the terms in the modified Hubbard model in Eq.~(1), (2) and (3). 
The fermions are moving in the potential 
$$
V_{\sigma, \mathrm{latt}}= V_{\sigma, x}\sin^2(\pi x/a) + V_{\sigma, y}\sin^2(\pi y/a) + V_{\sigma, z}\sin^2(\pi z/a),
$$
where $\sigma=\uparrow,\downarrow$ denotes the two species fermions and $V_{\sigma,x(y)(z)}$ are the corresponding lattice depths for $\sigma$-fermions along the $x,y,z$ direction
respectively. To create a two-dimensional ($2D$) geometry, we choose $V_{0}=V_{\downarrow, x}=V_{\downarrow, y}$,  $V_{1}=V_{\downarrow, z}=V_{\uparrow, x(y)(z)}$, 
and $V_1 \gg V_0$, which means that the $\downarrow$-fermions can effectively move in the $x-y$ plane with the $z$ motion frozen.
The contact-interaction Hamiltonian is given by,
\begin{equation} \label{Supeq1}
 H_{\rm con}=\frac{g}{2} \sum_{\sigma \neq \sigma'} \int \hat{\Psi}^{\dagger}_{\sigma}(\vec{r})\hat{\Psi}^{\dagger}_{\sigma'}(\vec{r})\hat{\Psi}_{\sigma'}(\vec{r})\hat{\Psi}_{\sigma}(\vec{r}) d\vec{r},
\end{equation}
where the field operators $\hat{\Psi}^{\dagger}_{\sigma}(\vec{r}), \hat{\Psi}_{\sigma}(\vec{r})$ denote the creation and destruction operators at position $\vec{r}$ for fermionic species 
$\sigma$. We also assume for simplicity that the mass of the two species is the same $m_{\uparrow}=m_{\downarrow}=m$. The contact interaction is given by $g=4\pi\hbar^2a_s/m$. 
From that, we construct the  Wannier functions  ${\cal W}^M_{\mathbf{i},\sigma}(x,y,z)=\omega^{m_x}_{i_x,\sigma}(x)\omega^{m_y}_{i_y,\sigma}(y)
\omega^{m_z}_{i_z,\sigma}(z)$ localized at the site $\mathbf{i}=(i_x,i_y,i_z)$, which correspond to the band $M=(m_x,m_y,m_z) $ \cite{Kohn}. Due to strong trapping along the $z$ direction, we only take into account
the lowest level in that direction. By expanding the field operators in the Wannier basis, we derive the parameters for the Hubbard model. In
particular, the integrals used to calculate the $s-p$ hopping term $J_{01}$ and the correlated hopping term in $p$-band $J_{11}$ are:
\begin{eqnarray}
\frac{J_{01}}{E_R} &=& \frac{8\pi^2a_s}{a}\int \mathrm{d}\boldsymbol{r} {\cal W}^{100}_{\mathbf{i},\downarrow}(\boldsymbol{r}) \left[{\cal W}^{000}_{\mathbf{i},\uparrow}(\boldsymbol{r})\right]^2
{\cal W}^{000}_{\mathbf{j},\downarrow}(\boldsymbol{r}), \nonumber\\
\frac{J_{11}}{E_R} &=& \frac{8\pi^2a_s}{a}\int \mathrm{d}\boldsymbol{r} {\cal W}^{100}_{\mathbf{i},\downarrow}(\boldsymbol{r}) \left[{\cal W}^{000}_{\mathbf{i},\uparrow}(\boldsymbol{r})\right]^2
{\cal W}^{100}_{\mathbf{j},\downarrow}(\boldsymbol{r}), \nonumber\\
\end{eqnarray}
where $\mathbf{ij}$ denote the nearest neighbouring sites along $x$-direction. As depicted in Fig.1(b)and (c) in the main paper, the effective tunneling in the $p$-band is given by
$J_p=J_1+2J_{11}$. Subsequently corresponding to the Hamiltonian in Eq. (2), we plot the magnitudes of the corresponding parameters in
Fig.\ref{Supfig1}. We see that with increasing attraction, the effective hybridized tunneling $J_{01}$ becomes comparable to the tunneling in the $p$-band, denoted by $J_p$.
Additionally we also plot the energy cost $\Delta$ (as defined after Eq. (2) in the main manuscript) as a function of effective interaction $a_s/a$ in Fig.\ref{Supfig1}.
For small $|a_s|/a_s$, the energy cost $\Delta$ is positive. As one increases the attraction, $\Delta$ decreases and for $a_s/a \lessapprox -0.56$, $\Delta$ becomes negative.
Now the appearance of {\it CH1} state is favoured when the $s-p$ tunneling strength becomes of the same order of magnitude as the tunneling in $p$-band with $J_{01}\approx J_p=J_1+J_{11}$ 
for $\Delta \lesssim 0$. From \ref{Supfig1} we see that around $ |\alpha| \sim 0.5-0.6$ both the tunnelings have the same order of magnitude facilitating the checkerboard phase.

\section{Derivation of $U_3$ and $U_2$ in the modified Hamiltonian}\label{intap}

Next we describe procedure to generate the effective interactions $U_2$ and $U_3$ in the modified Hamiltonian Eq.(2). As described in the paper, one of the  main parameters
which controls the transition is energy cost $\Delta=E_1-|U_3|+|U_2|$. Thus the main quantity to consider is the difference $U_3-U_2$. To do that first we expand \eqref{Supeq1}
in terms of the Wannier functions at the site $\mathbf{i}$,
\begin{eqnarray}
H_{\mathbf{i}}&=&\sum_{MNPQ}f_{MNPQ} \hat{c}^{\dagger}_{M,\mathbf{i}}\hat{b}^{\dagger}_{N,,\mathbf{i}}\hat{b}_{P,,\mathbf{i}}\hat{c}_{Q,,\mathbf{i}} \nonumber\\
&+& \sum_M E^c_M \hat{c}^{\dagger}_{M,\mathbf{i}}\hat{c}_{M,\mathbf{i}} + \sum_M E^b_M \hat{b}^{\dagger}_{M,\mathbf{i}}\hat{b}_{M,\mathbf{i}} \nonumber\\
\end{eqnarray}
where $MNPQ$ are the band indices and $\hat{c}^{\dagger}_{M,\mathbf{i}},\hat{c}_{M,\mathbf{i}}$ denote the creation and annihilation operators for the $\downarrow$-fermions at the site 
$\mathbf{i}$ and the band $M$. Similarly, $\hat{b}^{\dagger}_{N,\mathbf{i}}, \hat{b}_{N,\mathbf{i}}$ denote the creation and annihilation operators for the $\uparrow$-fermions at the site 
$\mathbf{i}$ and the band $N$. $E^c_M$ and $E^b_M$ are the single-particle energies for the $\downarrow$ and $\uparrow$-fermions respectively at the band $M$. 
The effective strengths $f_{NMPQ}$ are given in terms of Wannier functions as, 
$$
\frac{f_{MNPQ}}{E_R}=\frac{8\pi^2a_s}{a}\int \mathrm{d}\boldsymbol{r} {\cal W}^M_{\mathbf{i},\downarrow}(\boldsymbol{r}){\cal W}^N_{\mathbf{i},\uparrow}(\boldsymbol{r})
{\cal W}^P_{\mathbf{i},\uparrow}(\boldsymbol{r}) {\cal W}^Q_{\mathbf{i},\downarrow}(\boldsymbol{r}).
$$

Now to determine $U_2$, we first assume that the particles  occupy the lowest band. Then we calculate the effect of higher bands within the second order perturbation theory 
by taking into account transitions to higher bands. Then the Hamiltonian is,
\begin{widetext}
\begin{eqnarray}
H_{2} &=& -|f_{0000}|\hat{c}^{\dagger}_{0,\mathbf{i}}\hat{b}^{\dagger}_{0,\mathbf{i}}\hat{b}_{0,\mathbf{i}}\hat{c}_{0,\mathbf{i}} + \sum_M E^c_0 \hat{c}^{\dagger}_{0,\mathbf{i}}\hat{c}_{0,\mathbf{i}} + 
\sum_M E^b_M \hat{b}^{\dagger}_{0,\mathbf{i}}\hat{b}_{0,\mathbf{i}}, \nonumber\\
H_{2 \rm pert}&=& \sum_{M>0} f_{M000}\hat{c}^{\dagger}_{M,\mathbf{i}}\hat{b}^{\dagger}_{0,\mathbf{i}}\hat{b}_{0,\mathbf{i}}\hat{c}_{0,\mathbf{i}} +
\sum_{M>0N>0} f_{MN00}\hat{c}^{\dagger}_{M,\mathbf{i}}\hat{b}^{\dagger}_{M,\mathbf{i}}\hat{b}_{0,\mathbf{i}}\hat{c}_{0,\mathbf{i}}, \nonumber\\
&&
\end{eqnarray}
\end{widetext}
where in the diagonal term $H_2$, the first term is the interaction energy of the fermions in the lowest band and the next two terms denote the single-particle energies of the lowest bands 
for the $c$- and $b$-fermions. In the perturbative Hamiltonian $H_{2\rm pert}$, the first term denotes the transition of fermion species $c$-($=\downarrow$) to higher levels due to the 
interaction whereas the last term denotes the process where both $c$-($=\downarrow$) and $b$-($=\uparrow$)fermions are transferred to an excited state. Then for perturbation theory to be valid, the first condition is,
\begin{eqnarray}
|\frac{f_{M000}}{(E^c_M-E^c_0)+|f_{0000}|-|f_{M00M}|}| &\ll & 1, \nonumber\\
|\frac{f_{MN00}}{(E^c_M-E^c_0)+(E^b_N-E^b_0)+|f_{0000}|-|f_{MNNM}|}| &\ll& 1 \nonumber\\
\end{eqnarray}
To look into their properties, first we note that $|f_{MNNM}|, |f_{M00M}|, |f_{M000}|, |f_{MN00}| < |f_{0000}|$ as interaction in the lowest band has the strongest value. In addition,
$(E^c_M-E^c_0)>0, (E^b_N-E^b_0)>0$ for band indices $M,N>0$. So the denominators are always positive and we numerically checked that the fractions are much less than unity. 
This situation is drastically different for repulsive interactions where the denominator can indeed vanish making the perturbation theory invalid. Then within second-order perturbation 
theory we can write the two-fermion interaction energy,
\begin{widetext}
\begin{eqnarray} \label{two}
U_2 &=& -|f_{0000}|-\sum_{M>0} \frac{f^2_{M000}}{(E^a_M-E^a_0)+|f_{0000}|-|f_{M00M}|} - \sum_{MN>0} \frac{f^2_{MN00}}{(E^a_M-E^a_0)+(E^b_N-E^b_0)+|f_{0000}|-|f_{MNNM}|}
\end{eqnarray}
Similarly one can write the Hamiltonian pertaining to the situation when there are two $\downarrow$ ($c$-) particles, one at the $s$-band and another at the $p_x$-band, and one 
$\uparrow$ ($b$-) fermion in the $s$-band. The corresponding interaction energy $U_3$ is written in second-order perturbation as,
\begin{eqnarray} \label{three}
U_3 &=& -|f_{0000}| - |f_{1001}| - \sum_{M\ne [0,1]} \frac{f^2_{M000}}{(E^c_M-E^c_0)+|f_{0000}|-|f_{M00M}|} - \sum_{M\ne [0,1]} \frac{f^2_{M001}}{(E^c_M-E^c_1)+|f_{1001}|-|f_{M00M}|} \nonumber\\
&-& \sum_{M\ne [0,1] N>0} \frac{f^2_{MN00}}{(E^c_M-E^c_0)+(E^b_N-E^b_0)+|f_{0000}|+|f_{1001}|-|f_{1MM1}| - |f_{MNNM}|} \nonumber\\
&-& \sum_{M\ne [0,1] N>0} \frac{f^2_{MN01}}{(E^c_M-E^c_1)+(E^b_N-E^b_0)+|f_{0000}|+|f_{1001}|-|f_{0MM0}| - |f_{MNNM}|}, \nonumber\\
&&
\end{eqnarray}
\end{widetext}
where the band index $1=(100)$ denotes the $p_x$-band. The individual series in Eqs.\eqref{two}, \eqref{three} do not converge with respect to the summation over band indices $M,N$ 
and one needs to regularize the interaction at higher energies. But in this paper we are only interested in the difference in energy $U_3-U_2$ which converges as one takes bands with higher 
energies. In our parameter regime $U_3-U_2$ converges for band indices $M=15$. Convergence of the differences between the energies is also discussed in Ref.\cite{Tie} using the harmonic 
approximation for the lattice sites.

\section{Numerical methods} \label{simu}
To search for an optimal configuration of composites we use the simulated annealing method. This technique takes random walks through
the problem space and successively lower the temperature-like parameter. The probability of accepting a configuration is determined by the Boltzmann distribution what allows to get out of local minimum. We start our calculations from the phase separated configuration and configuration for each next step is chosen by randomly changing places of $n_c$ composites, where $n_c \le N_L/6$  for $N_L$ being the size of the lattice. Parameters of the calculations obviously depend on the lattice size. For $8x8$ lattice we have used the following: the initial temperature-like control parameter is lowered over time by use of a cooling schedule:
$T (t + 1) =T (t)/\mu_T$, where $\mu_T=1.008$; it starts at $T (0) = 0.009$ and continues until $T (t) < 1.0e-6$. For each step we try $n_{tries}=150$ configurations and for each temperature we perform $n_{iters}=200$ iterations. 
The parameters for $12x12$ ($16x16$) lattice are $\mu_T=1.002\ (1.001)$,  $n_{tries}=400 \ (600)$,  $n_{iters}=500 \ (800)$ iterations. Initial and final temperatures are the same for every lattice size. Simulated annealing gives us an approximate solution that with high probability is the global minimum. However it may happen that obtained configuration is a local minimum. Hence - to eliminate such solutions - we perform {\it second check}:  we group all the obtained configurations for different lattice depths and interaction strengths and we treat this set as a new problem space. The small size of this space allows us to individually compare the energies of every configurations.

\section{Effect of the tunneling of the $\uparrow$-fermions in deeper lattices}

In this section we study the effect of tunneling of the $\uparrow$-fermions on the Lieb lattice phase. Specially we are interested in
the case with $\Delta < 0$. The corresponding tunneling Hamiltonian for the $\uparrow$ fermions is written as,
\begin{equation}
H_{\uparrow t}=-J^1_{\uparrow} \sum_{\langle \mathbf{ij}\rangle} {\hat s_{\uparrow\mathbf{i}}}^{\dagger} + 
J^2_{\uparrow} \sum_{\langle \mathbf{ij}\rangle} {\hat s_{\uparrow\mathbf{i}}}^{\dagger}
\left(\hat{n}_{\downarrow i}+\hat{n}_{\downarrow j}\right)\hat{s}^{}_{\uparrow\mathbf{j}},
\end{equation}
where ${\hat s_{\mathbf{i}}}^{\dagger}, \hat{s}^{}_{\mathbf{i}}$ are the creation and annihilation operators for the $\uparrow$-fermions
at $s$-band and $J^1_{\uparrow}$ is the corresponding tunneling amplitude. The interaction-induced tunneling of the $\uparrow$-fermions in the 
$s$-band is denoted by $J^2_{\uparrow}$. When the band is filled for the $\downarrow$-fermions, then approximately each neighbour of a 
$\uparrow$-fermion is filled by a $\downarrow$-fermion. Then the total tunneling is given by,
$$
J_{\uparrow}=J^1_{\uparrow}+J^2_{\uparrow}.
$$
In the case of $\Delta < 0$ but small  the Lieb lattice structure is stable for sure provided $|J_{01}| \gg J_{\uparrow}$. This is the case in the strongly attractive limit as even
for $V_1=10E_R$ and $V_0=4E_R$, $|J_{01}|/J_{\uparrow} \sim 8$ with $a_s \sim -.6$. In the case when $\Delta \ll 0$, each composite occupied site is also occupied by a $\downarrow$-fermions in the $p$-orbital. Then the Lieb lattice structure is again stable 
provided $J_{01}/\Delta \gg J_{\uparrow}/E_1 $  which is also satisfied as the energy gap of the $p$-orbital ($E_1$) is much higher than $\Delta$ due to the attractive interaction. This condition can be proved trivially by looking into the second-order energy conserving processes which can delocalize the composite.


\begin{thebibliography}{10}
\bibitem{Lew2}
A review of such systems can be found on Chs. (6) and (8)  of \cite{Lew}.
\end{thebibliography}
\end{document}